\documentclass[12pt]{article}
\usepackage{sigsam}
\usepackage{amsmath, amssymb, url, caption, fancyhdr}

\usepackage{amsmath, amssymb, mathtools}
\usepackage{bbm}
\usepackage{enumerate} 
\usepackage{xcolor}

\usepackage[frozencache,cachedir=./_minted]{minted}
\setminted{
  fontsize=\small,
  breaklines,
  frame=single, 
  style=tango, 
}

\usepackage{xcolor}
\definecolor{myGreen}{rgb}{0.18039216 0.49803922 0.09411765}
\definecolor{darkred}{rgb}{0.65, 0.0, 0.0}

\usepackage[colorlinks,linkcolor=myGreen,citecolor=myGreen,urlcolor=darkred]{hyperref}
\usepackage[nameinlink,noabbrev]{cleveref}
\usepackage{thmtools}

\newcommand{\R}{\mathbb{R}}

\newcommand{\N}{\mathbb{N}}

\titlehead{SignatureTensors.jl: A Package for Signature Tensors in Julia}
\authorhead{ G.Riffo, L.Schmitz}
\issue{}
\articlehead{}

\begin{document}

\title{SignatureTensors.jl: A Package for Signature Tensors in Julia}
\author{Gabriel Riffo and  Leonard Schmitz \\
Institut für Mathematik, Technische Universität Berlin \\
10623 Berlin, Germany\\
\url{riffo@tu-berlin.de} , \url{lschmitz@math.tu-berlin.de}}
\date{}
\maketitle
\begin{abstract}
We introduce \texttt{SignatureTensors.jl}, a new package for computing signature tensors of paths in \texttt{julia}. We present its core functionality and demonstrate its use through illustrative examples. The package is compatible with the computer algebra system \texttt{OSCAR}, enabling both exact and numerical computations with signatures.
\end{abstract}

\section{Introduction}
Path signatures are fundamental objects in rough path theory \cite{friz2010multidimensional} and serve as a non-commutative feature that captures the essential geometry of sequential data. 
Recently, a tangible link to algebraic geometry \cite{bib:AFS2019} was established through the study of signature varieties associated with specific families of paths. 
This viewpoint proved particularly useful for studying the problem of learning paths from their signature tensors \cite{bib:PSS2019}. 

For this purpose, a practical and easily extendable package within a 
modern computer algebra system is required, providing access to multivariate arrays, Gröbner bases, Lie theory, 
non-commutative polynomials, and other structures. We introduce \texttt{SignatureTensors.jl}, a new package that leverages the symbolic computation capabilities of \texttt{OSCAR} \cite{OSCAR-book}, a modern open source computer algebra system written in \texttt{julia}.  
The package provides a general framework for computing and manipulating path signatures using algebraic methods while seamlessly interacting with the \texttt{OSCAR} ecosystem.

The previous approach \cite{RG20} based on the structural properties of free Lie algebras develops a \texttt{python} package to efficiently compute signatures of paths. Several frameworks for machine learning tasks are provided in \cite{diehl2024fruits,diehl2025tensortotensormodelsfastiterated}. Signature barycenters are efficiently computed using log-coordinates and symbolic preprocessing in \cite{clausel2024barycenterfreenilpotentlie}, based on BCH series and Gr\"obner bases of modules in \texttt{sage}. The recent \texttt{Macaulay2} package \cite{amendola2025computingpathsignaturevarieties} is close to our approach and implements several parts of \cite{bib:AFS2019}. Our package is compatible with the \texttt{OSCAR} code accompanying \cite{amendola2025learning,schmitz2025efficientalgorithmtensorlearning}. Our long-term goal is for \texttt{SignatureTensors.jl} to become a foundational tool for theoretical research on signature tensors, while providing a flexible base framework for interdisciplinary application, such as time series, spatial data, and more. The code and documentation are available at \url{https://leonardschmitz.github.io/SignatureTensors.jl}. The package is registered in the Julia General Registry and can be installed via \mintinline{julia}{Pkg.add("SignatureTensors")}.

%

\section{Path signatures}
We give a short introduction of our implementation of signatures. For more details on signatures in general, we refer to \cite{friz2010multidimensional}. 
For fixed dimension $d\geq 2$ and truncation level $k\in\N$ we consider the truncated tensor algebra
\begin{align*}T_{d,k}\;
:=\;\bigoplus\limits_{{\ell}={0}}^{k}(\R^{d})^{\otimes \ell}
&
\;=\;\R\oplus \R^{d}\oplus \R^{d\times d}\oplus\dots \oplus \R^{d\times \dots\times d}
\end{align*}
serving as our $\frac{d^{k+1}-1}{d-1}$ dimensional ambient space. 
 Let $X:[0,1]\rightarrow\R^d$ be a (smooth enough) continuous \emph{path} such that all integrals in \Cref{def:sig} below are defined. 
   The iterated-integrals signature
is a sequence of tensors 
$\sigma^{\leq k}(X):=1\oplus\sigma^{(1)}(X)\oplus\dots\oplus\sigma^{(k)}(X)\in T_{d,k}$  with  entries, 
\begin{align}\label{def:sig}
\sigma^{(\ell)}(X)_{w_1,\dots ,w_\ell}:= \int\limits_{0 \le t_1  \le \dots \le t_\ell \leq 1} \dot X_{w_1}(t_1) \dots \dot X_{w_\ell}(t_\ell)\,\mathrm dt_1 \dots \mathrm dt_\ell
\end{align}
for $1\leq w_1,\dots,w_\ell\leq d$.
For every path $X$, its signature $\sigma^{\leq k}(X)$ lies in the free nilpotent Lie group $\mathcal{G}_{d,k}$ of step $k$ over $\R^{d}$; see \cite[Theorem 7.30]{friz2010multidimensional}. We introduce two new structures \mintinline{julia}{TruncatedTensorAlgebra{R}} and \mintinline{julia}{TruncatedTensorAlgebraElem{R,E}} that implement the algebra of truncated tensor sequences $T_{d,k}$ and its elements, with a flexible type for its coefficients.
The constructors allow several base algebras, e.g., rational numbers or polynomial rings in \texttt{OSCAR}. We also support \texttt{julia} floats.

We implement (truncated) signatures of several path types, such as moment, axis, and polynomial paths, via the function \mintinline{julia}{sig}. Some of these path types require additional arguments. For example, the  option \mintinline{julia}{geom_type=:pwln} for piecewise linear paths requires the mandatory argument \mintinline{julia}{coef} that contains the coefficients for the linear segments. Here, the optional keyword \mintinline{julia}{algorithm::Symbol} specifies two algorithms: by default, we use
     \mintinline{julia}{algorithm=:Chen}, which iteratively applies Chen's identity \cite[Theorem 7.11]{friz2010multidimensional}, requiring $\mathcal{O}(m d^k)$ operations. If 
     \mintinline{julia}{algorithm=:congruence}, we use matrix--tensor congruence \cite[Equation~(3)]{bib:PSS2019} and require $\mathcal{O}(m^k + m d^k)$ elementary operations. We illustrate wall clock times for these options in \Cref{sec:benchmarks}. 
More generally, the option \mintinline{julia}{geom_type=:spline} implements piecewise polynomial paths with regularity constraints. Here we require the mandatory arguments \mintinline{julia}{coef::AbstractMatrix} and \mintinline{julia}{composition::Vector{Int}}, containing the coefficients using matrix--tensor congruence; see \cite[Corollary 4.11]{bib:AFS2019}. 



An important inverse problem is the task of recovering a path from its signature.
Recall from \cite{bib:AFS2019} that the image of the signature under piecewise linear paths with $m$ segments  
 is a semi-algebraic set in $T_{d,k}$. From  \cite{bib:PSS2019} we reduce the question whether an element $S$ is the signature of a piecewise linear path with $m$ segments to a polynomial system in $md$ variables with $(d^{k+1}-1)/(d-1)$ equations, each of degree $k$ or lower. For further details, we refer to \cite[Section 6]{amendola2025learning} with a notation closest to ours. Note that we learn with respect to the truncated sequence of signature tensors. With  \cite[Corollary 6.2]{bib:AFS2019} we obtain this sequence from its highest projection; see also \cite[Remark 3.5]{amendola2026signaturevarietiessplines} for details. 

In the following, we demonstrate learning within our package. 
This example shows that there are four (potentially complex) solutions corresponding to piecewise linear paths with four segments and the same signature as $S$.

\begin{minted}[escapeinside=||]{julia-repl} 
julia> using Oscar, SignatureTensors;
julia> A = QQ[ 6  -2  6   -10; 7  -4  10  -4];         # dimension d=2, m=4 segments 
julia> R, a = polynomial_ring(QQ, :a => (1:2, 1:4));       
julia> T = TruncatedTensorAlgebra(QQ,2,4);             # truncation level k=4
julia> S = sig(T,:pwln,coef=A);              # S signature of piecewise linear path
julia> aC = sig(T,:pwln,coef=a);             # aC signature with variable segments
julia> I = ideal(R,vec(S-aC));               # ideal generated by 31=2^5-1 relations
julia> dim(I), degree(I)                     # dimension and degree of I
(0, 4)
\end{minted}

If the dimension of the resulting vanishing  ideal is zero, and its degree one, then we have a unique path recovery. This has recently been formalized by the notion of \emph{path recovery degree}; see \cite[Definition 6.1]{amendola2026signaturevarietiessplines}. 
Then, the command \mintinline{julia}{recover} produces the coefficients of the piecewise linear path. The optional argument \mintinline{julia}{core} allows other core tensors. 
If $k=3$ and the coefficient matrix is invertible, then the solution is unique by \cite[Theorem 6.2]{bib:PSS2019}. 
With the option \mintinline{julia}{algorithm=:Sch25} we support the algorithm from \cite{schmitz2025efficientalgorithmtensorlearning} which requires $\mathcal{O}(d^4)$ elementary operations in expectation.

\section{Benchmarks}\label{sec:benchmarks}
Benchmarks were conducted on a MacBook Pro (Model Identifier: Mac17,2) equipped with an Apple M5 chip with 10 CPU cores (4 performance and 6 efficiency cores) and 24\,GB of RAM, running macOS 26.0 and Julia 1.12.3. The reported timings correspond to wall-clock times in milliseconds (rounded down). Timings were measured using \texttt{BenchmarkTools.jl}. For each parameter configuration we report the median wall-clock time across 100 samples. All experiments were carried out over \mintinline{julia}{base_algebra=Float64}. For path signatures and for each triple $(d,m,k)$, we generated random $d\times m$ integer matrices for \mintinline{julia}{coef} with entries sampled uniformly from $[-20,20]$. In \Cref{table:timingsForPwLin} we report on the wall-clock times for \mintinline{julia}{sig} with different algorithms. Further benchmarks are provided in the associated \href{https://leonardschmitz.github.io/SignatureTensors.jl}{repository}. 

\begin{table}[h]
\centering
\begin{minipage}{0.4\textwidth}
\centering
\begin{tiny}
\begin{tabular}{|c|c|c|c|c|c|c|}
\hline
$d\!\setminus\!m$  & 10 & 20 & 30 & 40 & 50 & 60 \\ \hline
10 & 0, 0 & 0, 0 & \textbf{0}, 1 & \textbf{0}, 2 & \textbf{0}, 5 & \textbf{1}, 9 \\ \hline
20 & 0, 0 & 0, 0 & \textbf{0}, 1 & \textbf{1}, 2 & \textbf{1}, 5 & \textbf{1}, 9 \\ \hline
30 & 0, 0 & 1, \textbf{0} & 2, \textbf{1} & 2, 2 & \textbf{3}, 5 & \textbf{4}, 9 \\ \hline
40 & 1, \textbf{0} & 2, \textbf{0} & 3, \textbf{1} & 5, \textbf{3} & 6, \textbf{5} & \textbf{7}, 9 \\ \hline
50 & 2, \textbf{0} & 5, \textbf{0} & 7, \textbf{1} & 10, \textbf{3} & 20, \textbf{5} & 39, \textbf{9} \\ \hline
60 & 4, \textbf{0} & 9, \textbf{0} & 21, \textbf{1} & 41, \textbf{3} & 55, \textbf{6} & 60, \textbf{10} \\ \hline
\end{tabular}
\caption*{$k=3$}
\end{tiny}
\end{minipage}
\hfill
\begin{minipage}{0.59\textwidth}
\centering
\begin{tiny}
\begin{tabular}{|c|c|c|c|c|c|c|}
\hline
$d\!\setminus\!m$  & 10 & 20 & 30 & 40 & 50 & 60 \\ \hline
10 & 0, 0 & \textbf{1}, 4 & \textbf{1}, 20 & \textbf{2}, 64 & \textbf{3}, 188 & \textbf{3}, 525 \\ \hline
20 & 4, \textbf{0} & 8, \textbf{4} & \textbf{20}, 21 & \textbf{40}, 76 & \textbf{49}, 189 & \textbf{57}, 541 \\ \hline
30 & 58, \textbf{1} & 243, \textbf{6} & 337, \textbf{23} & 372, \textbf{82} & 412, \textbf{200} & \textbf{458}, 558 \\ \hline
40 & 292, \textbf{3} & 360, \textbf{9} & 470, \textbf{31} & 726, \textbf{102} & 839, \textbf{219} & 1104, \textbf{566} \\ \hline
50 & 444, \textbf{11} & 913, \textbf{23} & 1307, \textbf{66} & 1742, \textbf{120} & 2127, \textbf{311} & 2437, \textbf{582} \\ \hline
60 & 1159, \textbf{39} & 2126, \textbf{67} & 2860, \textbf{96} & 4536, \textbf{153} & 7148, \textbf{434} & 8602, \textbf{636} \\ \hline
\end{tabular}
\caption*{$k=4$}
\end{tiny}
\end{minipage}
\caption{Timings (in milliseconds) for signatures of piecewise linear paths with $m$ segments using \mintinline{julia}{algorithm=:Chen} (first number) and \mintinline{julia}{algorithm=:congruence}   (second number).}
\label{table:timingsForPwLin}
\end{table}

Faster timings are highlighted in bold; Congruence outperforms Chen in high dimensions. Our timings are comparable with \cite{RG20}, e.g., the \texttt{numpy} library \texttt{iisignature} requires $6$ and $654$ milliseconds in the above setup for $d=m=60$ and $k\in\{3,4\}$. 
When moving to exact arithmetic over the rationals we use \texttt{OSCAR} and outperform the \texttt{M2} package \texttt{PathSignatures} from \cite{amendola2025computingpathsignaturevarieties}. For example, for $k=3$ and  $d=m\in\{10,20\}$ the package \texttt{PathSignatures} requires 1469 and 484595, whereas we need $1$ and $14$ miliseconds, respectively.

\subsection*{Acknowledgements}
We thank Carlos Am\'endola, Michael Joswig, Benjamin Lorenz, and Felix Lotter for several suggestions and comments on earlier versions of our package. The authors acknowledge funding by the Deutsche
Forschungsgemeinschaft (DFG, German Research Foundation) – CRC/TRR 388
``Rough Analysis, Stochastic Dynamics and Related Fields'' – Project A04 and B01, 
516748464.

\bibliographystyle{alpha} 
\bibliography{refs.bib}

\newcommand{\etalchar}[1]{$^{#1}$}
\begin{thebibliography}{CDM{\etalchar{+}}24}

\bibitem[AFS19]{bib:AFS2019}
Carlos Améndola, Peter Friz, and Bernd Sturmfels.
\newblock {Varieties of signature tensors}.
\newblock {\em Forum of Mathematics, Sigma}, 7, 2019.

\bibitem[ALS26]{amendola2026signaturevarietiessplines}
Carlos Améndola, Felix Lotter, and Leonard Schmitz.
\newblock Signature varieties of splines.
\newblock {\em preprint arXiv:2602.13011, to appear in Proceedings of the 2026
  International Symposium on Symbolic and Algebraic Computation}, 2026.

\bibitem[AS25]{amendola2025learning}
Carlos Am{\'e}ndola and Leonard Schmitz.
\newblock Learning barycenters from signature matrices.
\newblock {\em preprint arXiv:2509.07815, to appear in SIAM Journal on Matrix
  Analysis and Applications}, 2025.

\bibitem[ASLF26]{amendola2025computingpathsignaturevarieties}
Carlos Améndola, Angelo~El Saliby, Felix Lotter, and Oriol~Reig Fité.
\newblock {Computing Path Signature Varieties in Macaulay2}.
\newblock {\em Journal of Software for Algebra and Geometry}, 2026.

\bibitem[CDM{\etalchar{+}}24]{clausel2024barycenterfreenilpotentlie}
Marianne Clausel, Joscha Diehl, Raphael Mignot, Leonard Schmitz, Nozomi
  Sugiura, and Konstantin Usevich.
\newblock {The Barycenter in Free Nilpotent Lie Groups and Its Application to
  Iterated-Integrals Signatures}.
\newblock {\em SIAM Journal on Applied Algebra and Geometry}, 8(3):519--552,
  2024.

\bibitem[DEF{\etalchar{+}}24]{OSCAR-book}
Wolfram Decker, Christian Eder, Claus Fieker, Max Horn, and Michael Joswig,
  editors.
\newblock {\em The {C}omputer {A}lgebra {S}ystem {OSCAR}: {A}lgorithms and
  {E}xamples}, volume~32 of {\em Algorithms and {C}omputation in
  {M}athematics}.
\newblock Springer, 1 edition, 8 2024.

\bibitem[DISW26]{diehl2025tensortotensormodelsfastiterated}
Joscha Diehl, Rasheed Ibraheem, Leonard Schmitz, and Yue Wu.
\newblock Tensor-to-tensor models with fast iterated sum features.
\newblock {\em Neurocomputing}, 675:132884, 2026.

\bibitem[DK24]{diehl2024fruits}
Joscha Diehl and Richard Krieg.
\newblock {FRUITS: feature extraction using iterated sums for time series
  classification}.
\newblock {\em Data Mining and Knowledge Discovery}, 38(6):4122--4156, 2024.

\bibitem[FV10]{friz2010multidimensional}
Peter~K Friz and Nicolas~B Victoir.
\newblock {\em Multidimensional stochastic processes as rough paths: theory and
  applications}, volume 120.
\newblock Cambridge University Press, 2010.

\bibitem[PSS19]{bib:PSS2019}
Max Pfeffer, Anna Seigal, and Bernd Sturmfels.
\newblock {Learning Paths from Signature Tensors}.
\newblock {\em SIAM Journal on Matrix Analysis and Applications}, 2019.

\bibitem[RG20]{RG20}
Jeremy~F. Reizenstein and Benjamin Graham.
\newblock {Algorithm 1004: The Iisignature Library: Efficient Calculation of
  Iterated-Integral Signatures and Log Signatures}.
\newblock {\em ACM Trans. Math. Softw.}, 46(1), March 2020.

\bibitem[Sch25]{schmitz2025efficientalgorithmtensorlearning}
Leonard Schmitz.
\newblock An efficient algorithm for path recovery from signature tensors.
\newblock {\em preprint arXiv:2512.14218}, 2025.

\end{thebibliography}

\end{document}